\documentclass[12pt]{article}
\usepackage{amsmath}
\usepackage{amsthm}
\usepackage{amssymb}
\def\Arg{\operatorname{Arg}}

\def\Ima{\operatorname{Im}}
\begin{document}
\title{Entire functions, PT-symmetry
and Voros's quantization scheme}
\author{Alexandre Eremenko\thanks{Supported by NSF grant DMS-1361836.}
}
\date{September 11, 2020}
\maketitle
\begin{abstract}
In this paper, A. Avila's theorem
on convergence of the exact quantization scheme of A.~Voros 
is related to the reality proofs of eigenvalues of certain $PT$-symmetric
boundary value problems.
As a result, a special case
of a conjecture of C. Bender, S. Boettcher
and P. Meisinger on reality of eigenvalues is proved. 
\vspace{.1in}

AMS Class: 81Q05, 34M60, 34A05.

\end{abstract}

\noindent
{\bf 1. Introduction}. In this paper two theorems are proved:
\vspace{.1in}

\noindent
{\bf Theorem 1.}
{\em Consider three rays: 
$$L_j=\{ e^{ij\alpha}t:t\geq 0\},\quad j\in\{-1,0,1\},\quad i=\sqrt{-1}.$$
If
\begin{equation}\label{alpha}
\alpha\in(0,\pi/3],
\end{equation}
then there exists an entire function $g$ 
whose all zeros lie on $L_0$ and all $1$-points on $L_1\cup L_{-1}$,
and having infinitely many zeros and $1$-points.}
\vspace{.1in}

\noindent
{\bf Theorem 2.} {\em Consider the eigenvalue problem
\begin{equation}\label{ode}
-w''+(-1)^\ell(iz)^mw=\lambda w,
\end{equation}
where $m\geq 2$ is real, and $(iz)^m$ is the principal branch,
$(iz)^m>0$ when $z$ is on the negative imaginary ray,
with boundary conditions
\begin{equation}\label{y}
w(te^{i\beta})\to 0,\quad t\to\infty,
\end{equation}
where
$$\beta=-\pi/2\pm\frac{\ell+1}{m+2}\pi.$$
If $\ell=2$, and $m\geq 4$, then all eigenvalues are positive.
}
\vspace{.1in}

Theorem 2 it he simplest case of a conjecture of Bender, Boettcher and
Meisinger
\cite{BBM1,BBM2}. 
When $m=2$, $\ell=1$, the eigenvalue problem (\ref{ode}), (\ref{y})
is the harmonic oscillator. When $m=4$, $\ell=2$, it is the quartic
oscillator.
When $m$ is an integer, $m\geq 3$, and $\ell$ is an integer in $[1,m]$, 
Theorem~2 was proved by Shin \cite{Shin}. Notice that the case
$m=3,\ell=2$ is not covered by Theorem 2.
When $m\geq 2$ and $\ell=1$ positivity of eigenvalues was proved in
\cite{DDT}, section 6.2.

When $m$ is not an integer,
the bound for $m\geq 4$ in Theorem~2 seems to be
exact:
almost all eigenvalues are non-real
when $\ell=2$ and $m\in(2,3)\cup(3,4)$, according to the computation in
\cite[Figs. 14,15]{BBM1}. Here $m=3$ is an exceptional value, covered
by the theorem of Shin, when
all eigenvalues are real.

We recall that an eigenvalue problem for a differential operator is
called $PT$-symmetric if it is invariant
with respect to
the change of the independent variable $z\mapsto\sigma(z)=-\overline{z}$.
This means that the equation and the boundary conditions are invariant.
If each of the two boundary conditions is invariant under $\sigma$, the
problem is equivalent to an Hermitian one. 
In other $PT$-symmetric problems the two
boundary conditions are interchanged by $\sigma$.
$PT$-symmetric problems
have eigenvalues symmetric with respect to the real line
but not necessarily real.
The conjecture of Bender, Boettcher and Meisinger arises from
their numerical
study
of $PT$-symmetric boundary value problems for the operator (\ref{ode})
with various $PT$-symmetric boundary conditions. The idea was to connect
the potentials $z^2,z^3$ and $z^4$ into one continuous family. 
All our eigenvalue problems (\ref{ode}), (\ref{y}) are PT-symmetric.

The background of Theorem 1 and its relation with Theorem~2 is the following.

In a conference in Joensuu in summer 2015, Gary Gundersen
asked whether there exist entire functions with all zeros positive, while
$1$-points lie on some rays from the origin, 
distinct from the positive ray,
\cite[Questions 3.1, 3.2]{Gund}. As a partial answer to this question,
Bergweiler, Hinkkanen and the present author \cite{BEH}
proved among other things the
following fact:
\vspace{.1in}

\noindent
{\bf Theorem A.} {\em If there exists an entire function
with zeros on the positive ray $L_0$, and $1$-points on the rays $L$ and $L'$
from the origin, which are different from the positive ray,
and this function has infinitely many zeros
and $1$-points, then $\angle(L_0,L)= 
\angle(L_0,L')<\pi/2$.} 
\vspace{.1in}

Trying to construct an example of a function with this property, the authors
of \cite{BEH} recalled the functional equation
$$f(\omega \lambda)f(\omega^{-1}\lambda)=1-f(\lambda),
\quad \omega=e^{2\pi i/5},$$
which was studied by Sibuya and Cameron \cite{SC} and Sibuya \cite{Sibuya1}.
This equation is satisfied
by the Stokes multiplier of the differential equation
$$-y''+(z^3-\lambda)y=0.$$
On the other hand, it is known that this Stokes multiplier is an entire
function with all zeros positive \cite{DDT1}. So $f$ has positive zeros, and 
$1$-points of $f$ lie on the rays $\Arg z=\pm2\pi/5$.
Considering more general differential equations
\begin{equation}\label{de}
-y''+(z^m-\lambda)y=0,
\end{equation}
with integer $m\geq 3$
the authors of \cite{BEH} used the results of Sibuya \cite{Sibuya} and Shin
\cite{Shin} to prove Theorem 1 
with $\alpha=2\pi/(m+2)$, where $m\geq 3$ is an integer.

It was tempting to consider such differential equations (\ref{de}) with
non-integer $m\geq 2$, with solutions defined on the Riemann surface of the
logarithm. The Stokes multiplier of such an equation
is still an entire function of $\lambda$.
However, the numerical experiments and heuristic arguments
of Bender, Boettcher and Meisinger \cite{BBM1,BBM2} show that the straightforward generalization of the result of Shin on reality of $PT$-symmetric
eigenvalues does not hold for non-integer $m$.

This suggested a more general treatment of the required
functional equations (section 2 below) based on a deep result of Avila,
where the differential equation does not figure at all.
Theorem 2 is proved as a byproduct.

The main message of this paper is that a substantial
part of reality proofs for $PT$-symmetric eigenvalues
in \cite{DDT1,DDT,Shin} can be performed in a more general setting,
by working only with entire functions of the spectral
parameter $\lambda$, without
even mentioning the differential equation or the variable $z$.

A challenging question remains whether Theorem 1 can be extended
to angles $\alpha\in (\pi/3,\pi/2)$, besides $2\pi/5$.
Notice that Theorem 1 does not cover the case $\alpha=2\pi/5$ which was
proved in \cite{BEH}. Shin's proof of this result uses the change
of the independent variable $z\mapsto-z$ which in the case of equation
(\ref{ode}) works only for integer $m$.
Numerical and heuristic results in \cite{BBM1,BBM2} suggest
that the construction
described below will not work with
$\alpha\in(\pi/3,\pi/2)\backslash\{2\pi/5\}$.
\vspace{.1in}

{\em Remark.} Figs. 14, 15, 20 in \cite{BBM1}
show that for some non-integer $m\geq 2$ and some $\ell\geq 2$
almost all eigenvalues are non-real, and form complex conjugate pairs.
This shows that the usual asymptotic
expansions of eigenvalues $\lambda_k$ as a function of $k$,
which are common in one-dimensional eigenvalue problems
\cite{Shin2,Shin3,Fed},
and which would imply $|\lambda_{k+1}|>|\lambda_k|$, for large $k$,
cannot hold in these cases.
\vspace{.1in}

\noindent
{\bf 2. Voros's quantization scheme and Avila's theorem.}
\vspace{.1in}

Consider an entire function $f$ of genus zero with positive zeros and $f(0)=1$,
that is
\begin{equation}\label{f}
f(\lambda)=\prod_{j=1}^\infty\left(1-\frac{\lambda}{E_j}\right),
\quad 0<E_1<E_2\ldots.
\end{equation}
Denote
$$\omega=e^{i\alpha}, \quad \alpha\in(0,\pi/2).$$
Later, in section 3, we will need to impose a
stronger condition (\ref{alpha}).
Consider the function 
$$\Arg f(\omega^{-2}t)=\sum_{j=1}^\infty\tan^{-1}\frac{\sin2\alpha}{E_j/t-\cos2\alpha}.$$
This is a continuous, strictly increasing function of
$t$ which is zero at $0$, and tends to $+\infty$
as $t\to+\infty$. We want to find a function $f$ as in (\ref{f})
with the property
\begin{equation}\label{voros}
\frac{1}{\pi}\Arg f(\omega^{-2}E_k)=k-1/2,\quad k=1,2,3,\ldots.
\end{equation}
Avila proved in \cite{Avila} that such functions $f$ exist.
More precisely, Voros proposed to solve equations (\ref{voros})
in the following way. Start with an appropriate sequence $E=(E_k)$.
It determines
$f_E$ by (\ref{f}) and the increasing function
$t\mapsto \Arg f_E(\omega^{-2}t)$. Let $E'=(E_k^\prime)$ be
the solutions of
$$\frac{1}{\pi}\Arg f_E(\omega^{-2}E_k^\prime )=k-1/2,\quad k=1,2,\ldots.$$
These $E_k^\prime$ are uniquely defined because $t\mapsto\Arg f_E(\omega^{-2}t)$
is strictly increasing and maps $[0,+\infty)$ onto itself.
This construction defines a map $E\mapsto E'$.
Voros conjectured that under an appropriate choice of the initial
sequence 
iterates of this
map converge to a solution of (\ref{voros}).
This he called the ``exact quantization scheme''.
Avila proved the convergence of the scheme for every $\alpha\in (0,\pi/2)$.
(He uses parameter $\theta=\pi-2\alpha\in(0,\pi)$ instead of $\alpha$.)
The sufficient conditions of convergence
and initial conditions are stated on p.~309
in \cite{Avila}. In fact his assumptions on the right hand side of (\ref{voros})
are flexible: it has to be $k+O(1)$ and $\geq (k-1/2)(1-2\alpha)/\pi$.
\vspace{.1in}

\noindent
{\bf 3. Functional equations.}
\vspace{.1in}

It follows from (\ref{voros}) that the entire function
$$f(\omega^{-2}\lambda)+f(\omega^2 \lambda)$$
has zeros at $E_k$, and no other positive zeros. Indeed,
for $\lambda>0$ the summands are complex conjugate to each other,
so their sum is zero if and only if their arguments are $\pi/2$
modulo $\pi$, and this happens exactly for $\lambda=E_k$ according to
(\ref{voros}).
Therefore,
\begin{equation}\label{func1}
f(\omega^{-2}\lambda)+f(\omega^2 \lambda)=C(\lambda)f(\lambda),
\end{equation}
where $C$ is an entire function {\em without positive zeros}.
This is our first main functional
equation.

Equation (\ref{func1}) is equivalent to (\ref{voros}):
if $f$ is an entire function of the form (\ref{f}), satisfying (\ref{func1})
with some entire $C$ having no positive zeros, then $f$ satisfies (\ref{voros}).

Substituting $\lambda\mapsto\omega^2\lambda$ we obtain
\begin{equation}\label{func2}
f(\lambda)+f(\omega^4\lambda)=C(\omega^2\lambda)f(\omega^2\lambda).
\end{equation}
Elimination of $f(\lambda)$ from (\ref{func1}) and (\ref{func2})
gives
$$f(\omega^{-2}\lambda)=\left(C(\lambda)C(\omega^2\lambda)-1\right)f(\omega^2\lambda)
-C(\lambda)f(\omega^4\lambda).$$
By substituting $\lambda\mapsto\omega^{-1}\lambda$ and denoting
\begin{equation}\label{D}
D(\lambda)=C(\omega^{-1}\lambda)C(\omega \lambda)-1,
\end{equation}
we obtain our second main functional equation
\begin{equation}\label{func3}
f(\omega^{-3}\lambda)=D(\lambda)f(\omega \lambda)-C(\omega^{-1}\lambda)f(\omega^3\lambda),
\end{equation}
which is a direct consequence of (\ref{func1}).

Such functional equations were obtained first by Sibuya
\cite{Sibuya} in his studies of Stokes multipliers (the Stokes multiplier is $C$). Later it was discovered
by Dorey, Dunning and Tateo that the same functional equations occur in
the exactly solvable models of statistical mechanics
on two-dimensional lattices,
as well as in the quantum field theory \cite{DDT1}. 
Our new observation here is that all these
functional equations can be obtained from (\ref{voros}), without any
appeal 
to differential equations.

In the next proposition we will prove that zeros of $C$
and $D$ are negative. Setting $g(\lambda)=-D(-\lambda)$ we will obtain that
zeros of $g$ are positive while $1$-points, which are zeros of $C(\omega^{-1}\lambda)C(\omega\lambda)$
lie on $L_1\cup L_{-1}$ in view of (\ref{D}),
which will prove the Theorem 1.
\vspace{.1in}

\noindent
{\bf Proposition.} {\em Let $f$ be an entire function of order
less than $1$ of the form (\ref{f}),
and suppose that (\ref{func1}) is satisfied with some entire function $C$
which has no positive zeros. Then (\ref{func3}) is satisfied with
$D$ as in (\ref{D}) and all zeros of $C$ are negative.
If (\ref{alpha}) holds then all zeros of $D$ are negative as well.}
\vspace{.1in}


{\em Proof.}
First we prove that zeros of $C$ are real.
The idea of this comes from \cite{DDT1}, see also
\cite{DDT}. 
Suppose that $C(\lambda)=0$.
Then (\ref{func1}) implies 
\begin{equation}\label{ddt2}
|f(\omega^2\lambda)|=|f(\omega^{-2}\lambda)|.
\end{equation}
{}From the explicit form of $f$ in (\ref{f}) we see that
the function $\theta\mapsto|f(re^{i\theta})|$ is even, $2\pi$-periodic,
and strictly increasing on $(0,\pi)$. Therefore (\ref{ddt2}) can hold
only with real $\lambda$.

As $C$ has no positive zeros, they are all negative.
From (\ref{func1}) we obtain $C(0)=2$, so
\begin{equation}\label{C}
C(\lambda)=2\prod_{k=1}^\infty
\left(1+\lambda/\lambda_k\right),\quad \lambda_1<\lambda_2<\ldots;
\end{equation}
Now we prove that zeros of $D$ are real.
The following ingenious argument is due to K. Shin \cite[Thm. 11]{Shin},
but we slightly generalize
his result.

Let $D(\tau)=0$. As $D$ is real by (\ref{D}),
we also have $D(\overline{\tau})=0$,
so without loss of generality we choose
\begin{equation}\label{Ima}
\Ima\tau\geq 0.
\end{equation}
We claim that 
\begin{equation}\label{shin}
|C(\omega^{-1}\tau)|=1.
\end{equation}
For this we will need the assumption (\ref{alpha}). From (\ref{D}) we obtain
\begin{equation}\label{eq1}
|C(\omega^{-1}\tau)C(\omega\tau)|=1.
\end{equation}
Then, as $\Ima\tau\geq 0$ and $\Ima\omega>0$, we obtain
$$|C(\omega\tau)|=\prod_{k=1}^\infty\left|1+\omega\tau/\lambda_k\right|\leq
\prod_{k=1}^\infty\left|1+\omega^{-1}\tau/\lambda_k\right|=|C(\omega^{-1}\tau)|,$$
because
$$|1+\omega\zeta|=|\omega^{-1}+\zeta|\leq|\omega+\zeta|=
|1+\omega^{-1}\zeta|\quad\mbox{when}\quad\Ima\zeta\geq 0,\quad\Ima\omega>0.$$
Then (\ref{eq1}) gives 
$$|C(\omega^{-1}\tau)|\geq 1.$$
On the other hand, when we plug $\lambda=\tau$ to (\ref{func3}), we obtain
$$1\leq |C(\omega^{-1}\tau)|=\left|\frac{f(\omega^{-3}\tau)}{f(\omega^3\tau)}
\right|=\prod_{k=1}^\infty\left|\frac{\omega^3\lambda_k-\tau}{\omega^{-3}\lambda_k-\tau}\right|\leq 1,$$
where we used (\ref{Ima}) and $\Ima\omega^3\geq 0$, which follow
from (\ref{alpha}). This establishes
the claim (\ref{shin}).

Once (\ref{shin}) is known, we substitute $\lambda=\tau$ to (\ref{func3}) again,
and obtain
$$|f(\omega^{-3}\tau)|=|f(\omega^{3}\tau)|,$$
which is similar to (\ref{ddt2}), and implies that $\tau$ must be real,
in the same way as (\ref{ddt2}) implied that $\lambda$ was real.

It remains only to show that zeros of $D$ are negative. In view of (\ref{D}),
and (\ref{C})
we have for $x>0$
$$D(x)+1=C(\omega^{-1}x)C(\omega x)=4\prod_{k=1}^\infty
\left(1+2(x\cos\alpha)/\lambda_k+(x/\lambda_k)^2\right)>4,$$
so $D$ has no positive zeros.

This completes the proof of the Proposition.

Theorem 1 is an immediate consequence: take $g(\lambda)=-D(-\lambda)$.
\vspace{.1in}

{\em Remark.} For future references we state a slight generalization
of the proposition to which the same proof applies.
\vspace{.1in}

{\em Let $f$ be given by (\ref{f}) and suppose that we have
\begin{equation}\label{new}
kf(\omega^2\lambda)+k^{-1}f(\omega^{-2}\lambda)=C(\lambda)f(\lambda),
\end{equation}
where $|k|=1$, and $C$ is an entire function with no positive zeros,
and $\omega=e^{i\alpha}$, where $0<\alpha\leq\pi/3$.
Then
\begin{equation}\label{main5}
k^{-3/2}f(\omega^{-3}\lambda)+C(\omega^{-1}\lambda)k^{3/2}f(\omega^3\lambda)=
D(\lambda)k^{1/2}f(\omega\lambda),
\end{equation}
where $D$ is as in (\ref{D}) and both $D$ and $C$ have all zeros negative.}

The proof is the same as for $k=1$.

Combining (\ref{main5}) with (\ref{new}) we can eliminate $C$ and express
$D$ directly in terms of $f$:
\begin{eqnarray}\label{main6}
&D(\lambda)f(\omega^{-1}\lambda)f(\omega\lambda)=\nonumber \\
&k^{-2}f(\omega^{-3}\lambda)f(\omega^{-1}\lambda)+
k^2f(\omega^3\lambda)f(\omega\lambda)+
f(\omega^{-3}\lambda)f(\omega^3\lambda).
\end{eqnarray}
\vspace{.1in}

\noindent
{\bf 4. Proof of Theorem 2.}
\vspace{.1in}

It is convenient to make the change of the variable
$y(z)=w(-iz)$. Then
\begin{equation}\label{ode2}
-y''+\left((-1)^{\ell+1}z^m+\lambda\right)y=0,
\end{equation}
and
\begin{equation}\label{y2}
y(z)\to 0,\quad z\to\infty,\quad
\arg z=\pm\frac{\ell+1}{m+2}\pi.
\end{equation}
In the equation (\ref{ode2}) the principal branch of $z^m$ is used,
so the branch cut is on the negative ray.

According to Sibuya \cite{Sibuya}, there is a unique normalized solution
$y_0(z,\lambda)$ of the equation (\ref{ode2}) with $\ell=1$ with the property
\begin{equation}\label{as1}
y_0(z,\lambda)=(1+o(1))z^{-m/4}\exp\left(-\frac{2}{m+2}z^{(m+2)/2}\right),
\end{equation}
as $z=te^{i\theta},\; t>0,\; t\to\infty$ and $|\theta|<3\pi/(m+2)$.
Moreover, for every fixed $z_0$, the function $y(z_0,\lambda)$ is
an entire function of $\lambda$ of order $1/2+1/m<1$.
Sibuya stated this result only for integer $m$, but his proof actually does
not depend on this assumption, see \cite{T}, \cite{DDT}, \cite{DDT2}.
Let
\begin{equation}\label{omega}
\omega=\exp(2\pi i/(m+2)).
\end{equation}
As $m\geq 2$, $\Arg\omega\in(0,\pi/2)$.
Then
$$y_k(z,\lambda)=\omega^{k/2}y_0(\omega^{-k}z,\omega^{2k}\lambda),$$
where $\omega^{k/2}:=\exp(\pi ik/(m+2)),$
satisfies the same differential equation (\ref{ode2}) with $\ell=1$
when $k$ is an integer,
and the equation (\ref{ode2}) 
with $\ell=2$ when $k$ is a half of an odd integer.
We use normalization of $y_k$
from \cite{DDT1,DDT} which is more convenient than Sibuya's normalization.
Any three solutions of the same
differential equation must be linearly dependent, so
$$y_1(z,\lambda)=C_0(\lambda)y_0(z,\lambda)
-\tilde{C}(\lambda)y_{-1}(z,\lambda).$$
Comparison of the asymptotics of $y_1$ and $y_{-1}$ gives $\tilde{C}\equiv 1$,
so 
\begin{equation}\label{main1}
y_1(z,\lambda)=C_0(\lambda)y_0(z,\lambda)-
y_{-1}(z,\lambda).
\end{equation}
One can show that $C_0$ is an entire function
of order $1/2+1/m$, \cite{Sibuya,T}.
Substituting $(z,\lambda)\mapsto (\omega^{-1}z,\omega^{2}\lambda)$ into 
(\ref{main1}), we obtain
\begin{equation}\label{main2}
y_0(z,\lambda)=C_0(\omega^2\lambda)
y_1(z,\lambda)-y_2(z,\lambda),
\end{equation}
a relation of the form (\ref{new}).
Eliminating $y_0(z,\lambda)$ from (\ref{main1}), (\ref{main2}), we obtain
$$y_{-1}(z,\lambda)=\left(C_0(\lambda)C_0(\omega^2\lambda)-1\right)
y_1(z,\lambda)-C_0(\lambda)y_2(z,\lambda)$$
Finally, substitute $(z,\lambda)\mapsto(\omega^{1/2}z,\omega^{-1}\lambda)$
and multiply on $\omega^{-1/4}$.
The result is
\begin{equation}\label{main3}
y_{-3/2}(z,\lambda)=D_0(\lambda)y_{1/2}(z,\lambda)-C_0(\omega^{-1}\lambda)y_{3/2}(z,\lambda),
\end{equation}
where
\begin{equation}\label{D0}
D_0(\lambda)=C_0(\omega^{-1}\lambda)C_0(\omega\lambda)-1.
\end{equation}
Equation (\ref{main3}) is a special case of (\ref{main5}).

We see that functions $y_{3/2}$ and $y_{-3/2}$ satisfy equation (\ref{ode2})
with $\ell=2$,
and tend to zero on the rays $\Arg z=3\pi/(m+2)$ and $\Arg z=-3\pi/(m+2)$,
respectively.
These functions, as functions of $z$, are linearly dependent
if and only if  
(\ref{ode2}), (\ref{y2}) with $\ell=2$ have a non-trivial
solution. 
Thus the eigenvalues for $\ell=2$ are
zeros of $D_0$.

Let us denote $f(\lambda)=\lim_{x\to 0+}y_0(x,\lambda)$;
it is easy to see that this
is well defined, despite the singularity of (\ref{ode2}) at $0$.
Then $f$ is an entire function of genus $0$ and its zeros $E_k$ can be
interpreted as eigenvalues of (\ref{ode2}) under the boundary conditions
\begin{equation}\label{br}
\lim_{t\to 0+}y(t)=\lim_{t\to+\infty}y(t)=0.
\end{equation}
This problem is self-adjoint, so all eigenvalues are real. Moreover
the potential $z^m$ is positive on the positive ray, so the ``eigenvalues''
$\lambda$ in (\ref{ode2}) with $\ell=1$ under the conditions
(\ref{br}) are all {\em negative}.
We also notice that $y_0(x,\lambda)$ is real for real $x$ and $\lambda$,
so $f(0)$ is real, and thus $f$ is a real entire function.

Plugging $z=0$ in (\ref{main1}) we obtain
\begin{equation}\label{ma1}
\omega^{-1/2}f(\omega^{-2}\lambda)+\omega^{1/2} f(\omega^2\lambda)=C_0(\lambda) f(\lambda)
\end{equation}
which is analogous to (\ref{new}). Notice that
$C_0(0)=\omega^{-1/2}+\omega^{1/2}$ is real.
All zeros of $C_0$ are positive by the results in \cite[6.2]{DDT} and
\cite{Shin4}.
Application of the Remark in the previous section gives that all
zeros of $D_0$ are negative, and this completes the
proof of Theorem~2.
%

The author thanks Andr\'e Voros and Kwang Shin for useful discussions.

{\em Department of Mathematics, Purdue University

West Lafayette, IN 47907 USA.}
\end{document}